%% file: main.tex
\documentclass[conference]{IEEEtran}
\IEEEoverridecommandlockouts
\usepackage{cite}
\usepackage{amsmath,amssymb,amsfonts}
\usepackage{algorithmic}
\usepackage{graphicx}
\usepackage{textcomp}
\usepackage{xcolor}
\usepackage{algorithm}      % for writing algorithms
\usepackage[utf8]{inputenc}
\usepackage{todonotes}
\usepackage{pgfplots}
\usepackage[acronym, shortcuts]{glossaries}
\usepackage{xargs}   % for multi-input-argument newcommands
\usepackage{macros}
\usepackage{orcidlink}  % for ORCiD badges/hyperlinks
\loadglsentries[type]{acro.tex}

\DeclareUnicodeCharacter{2212}{−}
\usepgfplotslibrary{groupplots,dateplot}
\usetikzlibrary{patterns,shapes.arrows,external}
\pgfplotsset{compat=newest}

\def\BibTeX{{\rm B\kern-.05em{\sc i\kern-.025em b}\kern-.08em
    T\kern-.1667em\lower.7ex\hbox{E}\kern-.125emX}}
\begin{document}

\title{
    % Improved Topology-Independent Distributed Adaptive Node-Specific Signal Estimation in Wireless Acoustic Sensor Networks\\
    Improved Topology-Independent Distributed Adaptive Node-Specific Signal Estimation for Wireless Acoustic Sensor Networks\\
\thanks{
    This research was carried out at the ESAT Laboratory of KU Leuven, in the frame of Research Council KU Leuven C14-21-0075 ``A holistic approach to the design of integrated and distributed digital signal processing algorithms for audio and speech communication devices'', and was supported by the European Union's Horizon 2020 research and innovation programme under the Marie Skłodowska-Curie grant agreement No. 956369: `Service-Oriented Ubiquitous Network-Driven Sound — SOUNDS'. The scientific responsibility is assumed by its authors. This paper reflects only the authors' views and the Union is not liable for any use that may be made of the contained information. 
}
}

% \author{
%     \IEEEauthorblockN{\textsuperscript{1}Paul Didier, \textsuperscript{1}Toon van Waterschoot, \textsuperscript{2,3}Simon Doclo, \textsuperscript{3}Jörg Bitzer, \textsuperscript{1}Marc Moonen}
%     \IEEEauthorblockA{
%         \textsuperscript{1}\textit{Dept. of Electrical Engineering (ESAT), STADIUS, KU Leuven}, Leuven, Belgium\\
%         \textsuperscript{2}\textit{Carl von Ossietzky Universität Oldenburg}, Oldenburg, Germany\\
%         \textsuperscript{3}\textit{Fraunhofer IDMT, Project Group Hearing, Speech and Audio Technology}, Oldenburg, Germany
%     }
% }

\author{
    \IEEEauthorblockN{Paul Didier\IEEEauthorrefmark{1} \orcidlink{0000-0002-1209-2614}, Toon van Waterschoot\IEEEauthorrefmark{1}\orcidlink{0000-0002-6323-7350}, Simon Doclo\IEEEauthorrefmark{2,3}\orcidlink{0000-0002-3392-2381}, Jörg Bitzer\IEEEauthorrefmark{3}\orcidlink{0000-0002-0595-0262}, Marc Moonen\IEEEauthorrefmark{1} \orcidlink{0000-0003-4461-0073}}
    \IEEEauthorblockA{\IEEEauthorrefmark{1}\textit{Dept. of Electrical Engineering (ESAT), STADIUS, KU Leuven}, Leuven, Belgium}
    \IEEEauthorblockA{\IEEEauthorrefmark{2}\textit{Carl von Ossietzky Universität Oldenburg}, Oldenburg, Germany}
    \IEEEauthorblockA{\IEEEauthorrefmark{3}\textit{Fraunhofer IDMT, Project Group Hearing, Speech and Audio Technology}, Oldenburg, Germany}
}

\maketitle

\begin{abstract}
    This paper addresses the challenge of topology-independent (TI) distributed adaptive node-specific signal estimation (DANSE) in wireless acoustic sensor networks (WASNs) where sensor nodes exchange only fused versions of their local signals. An algorithm named TI-DANSE has previously been presented to handle non-fully connected WASNs. However, its slow iterative convergence towards the optimal solution limits its applicability. To address this, we propose in this paper the TI-DANSE$^+$ algorithm. At each iteration in TI-DANSE$^+$, the node set to update its local parameters is allowed to exploit each individual partial in-network sums transmitted by its neighbors in its local estimation problem, increasing the available degrees of freedom and accelerating convergence with respect to TI-DANSE. Additionally, a tree-pruning strategy is proposed to further increase convergence speed. TI-DANSE$^+$ converges as fast as the DANSE algorithm in fully connected WASNs while reducing transmit power usage. The convergence properties of TI-DANSE$^+$ are demonstrated in numerical simulations.
\end{abstract}


\section{Introduction}

The rapid development of the Internet of Things has led to a ubiquity of acoustic sensors (microphones) in our environment. This enables the creation of \glspl*{wasn} where devices such as laptops, smartphones, or hearing aids~\cite{bertrandApplications,aliasReview} (referred to as ``nodes'' in this context) are wirelessly connected. %A \gls*{wasn} can cover a wider spatial area than centralized solutions, as the latter rely on fusion centers.
% Nodes in a \gls*{wasn} can thus obtain access to information about distant sound sources.
% By allowing each device to perform computations and exchange signals with other devices, it is possible to attain the same performance as an equivalent centralized system without a fusion center.
Research interest has grown over recent years towards the development of distributed algorithms for deployment in \glspl*{wasn}, aimed at solving various signal processing tasks with a performance as good as equivalent centralized systems~\cite{wuReview,cobosSurvey}. In this paper, the focus is set towards the task of node-specific signal estimation.

Distributed algorithms must meet the requirements of distributed systems, including transmit power constraints and limited communication bandwidth. To address these aspects, the \gls*{danse} algorithm has been developed~\cite{bertrandDistributed}. It allows nodes to iteratively reach the centralized signal estimation performance by exchanging low-dimensional (so-called ``fused'') versions of their sensor signals. Its application is, however, limited to \gls*{fc} \glspl*{wasn} where every node can communicate with every other node. Later, the topology-independent \gls*{danse} algorithm (\glsunset{tidanse}\gls*{tidanse}) has been proposed~\cite{szurleyTopology} for non-\gls*{fc} \glspl*{wasn} with a possibly time-varying topology.

An important drawback of \gls*{tidanse} is its slow convergence, which limits its applicability. This slow convergence occurs because, at each algorithmic iteration, the node set to update its local parameters (i.e., the updating node) only has access to the sum of all fused signals in the \gls*{wasn}. Compared to \gls*{danse} in \gls*{fc} \glspl*{wasn}, this significantly reduces the number of degrees of freedom (\glsunset{dof}\glspl*{dof}) available at the updating node to solve its local signal estimation problem.

In this paper, we address the slow convergence of \gls*{tidanse} by proposing \glsunset{tidansep}\gls*{tidansep}, an algorithm that allows the updating node to \textit{separately} exploit partial in-network sums of fused signals coming from its neighbors while solving its local signal estimation problem. Furthermore, we show that the convergence speed benefits of \gls*{tidansep} can be enhanced by introducing a particular tree-pruning strategy. In \gls*{fc} \glspl*{wasn}, \gls*{tidansep} converges as fast as \gls*{danse}, yet at a lower transmit power cost since \gls*{tidansep} requires fewer signals to be exchanged across the \gls*{wasn}.%functions based on peer-to-peer signal exchanges instead of broadcasting.

The organization of this paper is as follows. The context of the problem is given in~\secref{sec:problemStatement}, defining the signal model and the centralized solution, and reviewing the \gls*{danse} algorithm and the \gls*{tidanse} algorithm. The \gls{tidansep} algorithm is introduced in~\secref{sec:TIDANSEplus}. The proposed tree-pruning strategy is presented in~\secref{sec:treepruning}. Numerical experiments results are shown and discussed in~\secref{sec:numericalExperiments}. Finally, conclusions are formulated in~\secref{sec:conclusion}.

%%%%%
\section{Problem Statement and Related Algorithms}\label{sec:problemStatement}

Consider a \gls{wasn} consisting of $K$ nodes where any node $q\in\K=\{1,\dots,K\}$ has $M_q$ sensors (i.e., microphones). The total number of sensors is $M=\sum_{q\in\K}M_q$. 
All signals are assumed to be complex-valued to allow for frequency-domain representations. The signal captured at time $t$ by the $m$-th sensor of node $q$ is modelled as $y_{q,m}[t] =  s_{q,m}[t] + n_{q,m}[t]$ where $s_{q,m}[t]$ and $n_{q,m}[t]$ represent the desired and noise components, respectively. %The noise component $n_{k,m}[t]$ may contain contributions from localized or diffuse noise sources as well as thermal microphone noise.
For conciseness, the index $[t]$ is dropped in the following.
All sensor signals of node $q$ can be stacked in $\yk[q] =  [y_{q,1}\dots y_{q,M_q}]^\T\in\C[M_q]$, resulting in:

\begin{equation}\label{eq:signal_model}
    \yk[q] =  \sk[q] + \nk[q] = \Psibk[q]\mathbf{s}^\mathrm{lat} + \nk[q],
\end{equation}

\noindent
where $\sk[q]$ and $\nk[q]$ are defined similarly to $\yk[q]$, $\Psibk[q]\in\C[M_q][Q]$ is the steering matrix for node $q$, and $\mathbf{s}^\mathrm{lat}\in\C[Q]$ is the latent signal produced by $Q$ desired signal sources.

We consider a scenario where each node $q\in\K$ aims at estimating a $Q$-dimensional target signal $\dk[q] = \Ekk[q]^\T\sk[q] = \Psibkov[q]\mathbf{s}^\mathrm{lat}$ where $\Ekk[q]\in\{0,1\}^{M_q\times Q}$ is a selection matrix extracting $\dk[q]$ from $\sk[q]$ and $\Psibkov[q] = \Ekk[q]^\T\Psibk[q]$. Cases where the number of latent sources is different from the dimension of $\dk[q]$ are not considered here for simplicity.

%%%%
\subsection{Centralized Solution}

In a centralized scenario, each node has access to the centralized signal vector $\yk[] = [\yk[1]^\T\:\dots\: \yk[K]^\T]^\T\in\C[M]$. As in~\eqref{eq:signal_model}, $\yk[]$ can be separated into a desired signal component $\mathbf{s}$ and noise component $\mathbf{n}$. The following \gls*{lmmse} problem is then considered at any node $q\in\K$ to obtain the centralized optimal filter $\hWk[q]$, which is a \gls*{mwf}~\cite{golubMatrix}:

\vspace{-1em}
\begin{align}\label{eq:lmmseCentr}
    \hWk[q] &= \underset{\Wk[q]}{\mathrm{arg\,min}}\:
    \E[{
        \left\|
            \dk[q] - \Wk[q]^\Her\mathbf{y}
        \right\|^2_2
    }]\\
    &=\left(\Ryy\right)^{-1}\Rydk[q]%\label{eq:mwfCentr_prev}
    =\left(\Ryy\right)^{-1}\Rss\Ek[q],\label{eq:mwfCentr}
\end{align}

\noindent
with the selection matrix $\Ek[q]\in\{0,1\}^{M\times Q}$ extracting $\dk[q]$ from $\mathbf{s}$ and the spatial covariance matrices (\glsunset{scm}\glspl*{scm}) $\Ryy= \mathbb{E}\{\mathbf{yy}^\Her\}$, $\Rydk[q] = \mathbb{E}\{\mathbf{yd}_q^\Her\}$, and $\Rss = \E[{\mathbf{ss}^\Her}]$, where $\E[{\cdot}]$ is the expectation operator. Equation~\eqref{eq:mwfCentr} holds assuming that $\mathbf{s}$ and $\mathbf{n}$ are uncorrelated.
The estimation of $\Ryy$ and $\Rss$ can be achieved in practice through, e.g., time-averaging and source activity detection techniques (see, e.g.,~\cite{zhaoModel}) considering that $\Rss=\Ryy-\Rnn$ when $\mathbf{s}$ and $\mathbf{n}$ are uncorrelated. %More details on \gls*{scm} estimation can be found in~\cite{bertrandDistributed}.
% Estimating $\Ryy$ can be done by averaging over the available observations of $\mathbf{y}$, assuming short-term stationarity of the signals. Although $\dk$ is not directly available, $\Rydk$ can be estimated, assuming no correlation between desired and noise components, as:

It can be shown that, since $\dk[q] = \Psibkov[q]\mathbf{s}^\mathrm{lat}\fa q\in\K$, the centralized \gls*{mwf} solution~\eqref{eq:mwfCentr} at any two nodes $q$ and $q'$ are the same up to a $Q\times Q$ transformation since:

\vspace{-1em}

\begin{equation}\label{eq:centr_same_sol_space}
    \hWk[q] = \hWk[q']\Psibkq[q][q']
    \:\:
    \text{where}\:\:
    \Psibkq[q][q'] = \left(\Psibkov[q']^\Her\right)^{-1}\Psibkov[q]^\Her.
\end{equation}

% \begin{align}\label{eq:Rydk_RssEk}
%     \Rydk
%     &= \E[\mathbf{y}\dk^\Her]
%     = \E[\mathbf{s}\dk^\Her]\\
%     &= \E[\mathbf{ss}^\Her]\Ek
%     =  \Rss\Ek\label{eq:Rydk_RssEk_final}\\
%     &= \Ryy - \Rnn.
% \end{align}

% \noindent
% where the noise \gls*{scm} $\Rnn =  \mathbb{E}\{\mathbf{nn}^\Her\}$ can, e.g., be estimated by making use of the on-off characteristics of the desired speech signal and a \gls*{vad} or by using prior knowledge~\cite{zhaoModel, kimVoice}.
% In practical online-processing applications, the \glspl*{scm} may be estimated using, e.g., exponential averaging as:

% \begin{equation}\label{eq:expavg}
%     \begin{split}
%         \Ryy[t] &= \beta\Ryy[t-1] + (1-\beta)\mathbf{y}[t]\mathbf{y}[t]^\Her\:\:\text{if VAD$[t]=1$},\\
%         \Rnn[t] &= \beta\Rnn[t-1] + (1-\beta)\mathbf{y}[t]\mathbf{y}[t]^\Her\:\:\text{if VAD$[t]=0$},
%     \end{split}
% \end{equation}

% \noindent
% where $t$ is the time-frame index, $0\ll\beta<1$ is the forgetting factor, and VAD$[t]$ denotes the \gls*{vad} value at time-frame $t$.

%%%%%%
% \section{\gls*{danse} and \gls*{tidanse}}\label{sec:ti_danse}

%%%%
\subsection{\gls*{danse} and \gls*{tidanse}}\label{subsec:danse_and_tidanse}

To build $\yk[]$ in a distributed setting, every node should transmit all their sensor signals to every other node, consuming a substantial amount of transmit power. This issue can be addressed through the \gls*{danse} or \gls*{tidanse} algorithm, depending on the \gls*{wasn} topology. Indeed, in both \gls*{danse} and \gls*{tidanse}, nodes exchange low-dimensional fused versions of the multichannel local sensor signals, relieving part of the transmit power burden. The specifications of the two algorithms in sequential node-updating schemes~\cite{bertrandDistributed,szurleyTopology} are summarized below.
% Simultaneous or asynchronous node-updating schemes are not considered in this paper.

At each iteration $i\in\N$, each node $q\in\K$ defines a fusion matrix $\Pk[q]^i\in\C[M_q][Q]$ and computes a $Q$-channel fused signal $\zkd[q]^i = \Pk[q]^{i\Her}\yk[q]$. 
In practice, iterations are spread over time such that each $\zkd[q]^i$ is computed using a different frame $\yk[q]$.

In \gls*{danse}, i.e., in a \gls*{fc} \gls*{wasn}, the fused signals can be transmitted such that each node $q\in\K$ has access to the observation vector $\ty[q,\mathrm{D}]^{i} = \begin{bmatrix}
    \yk[q]^\T \: \zkd[1]^{i\T} \: \dots \: \zkd[q-1]^{i\T} \: \zkd[q+1]^{i\T} \: \dots \: \zkd[K]^{i\T}
\end{bmatrix}^\T= \begin{bmatrix}
    \yk[q]^\T \:|\: \zkd[-q]^{i\T}
\end{bmatrix}^\T\in\C[{(M_q+Q(K-1))}]$. The subscript ``D'' is used to unambiguously refer to the \gls*{danse} algorithm.

In a non-\gls*{fc} \gls*{wasn}, fused signals cannot be directly transmitted to every node. Instead, the \gls*{tidanse} algorithm can be used by pruning the \gls{wasn} to a tree at the beginning of $i$. All fused signals are then summed through a sequence of in-network partial summations from leaf nodes to the root node. We denote $\Kmk[q]$ as the set of all nodes but $q$, i.e., $\Kmk[q]=\K\backslash\{q\}$. Once $\ett^i = \sum_{q\in\K}\zkd[q]^i\in\C[Q]$ is available at the root node, it is propagated back through the entire \gls*{wasn}, such that each node $q\in\K$ accesses the observation vector $\ty[q,\mathrm{TI}]^i = [\yk[q]^\T\:|\: \ettmk[q]^{i\T}]^\T\in\C[{(M_q+Q)}]$ where $\ettmk[q]^{i} = \ett^{i} - \zkd[q]^i = \sum_{m\in\Kmk[q]}\zkd[m]^i\in\C[Q]$. The subscript ``TI'' refers to the \gls*{tidanse} algorithm.

In both \gls*{danse} and \gls*{tidanse}, the updating node $k$ then solves its own \gls*{lmmse} problem to estimate $\dk$ from the observation vector, i.e., $\ty[k,\mathrm{D}]^i$ as in~\eqref{eq:lmmseDANSE} or $\ty[k,\mathrm{TI}]^i$ as in~\eqref{eq:lmmseTIDANSE}:

\begin{align}\label{eq:lmmseDANSE}
    \begin{bmatrix}
        \Wk[kk,\mathrm{D}]^{i+1}\\
        \Gk[-k]^{i+1}
    \end{bmatrix}
    &=
    \underset{\Wkk,\Gk[-k]}{\mathrm{arg\,min}}\:
    \mathbb{E}\left\{
        \left\|
            \dk - \begin{bmatrix}
                \Wkk^\Her \:|\: \Gk[-k]^\Her
            \end{bmatrix}\begin{bmatrix}
                \yk\\
                \zkd[-k]^i
            \end{bmatrix}
        \right\|^2_2
    \right\}\\
    \begin{bmatrix}
        \Wk[kk,\mathrm{TI}]^{i+1}\\
        \Gk[k]^{i+1}
    \end{bmatrix}
    &=
    \underset{\Wkk,\Gk}{\mathrm{arg\,min}}\:
    \mathbb{E}\left\{
        \left\|
            \dk - \begin{bmatrix}
                \Wkk^\Her \:|\: \Gk^\Her
            \end{bmatrix}\begin{bmatrix}
                \yk\\
                \ettmk^i
            \end{bmatrix}
        \right\|^2_2
    \right\}.\label{eq:lmmseTIDANSE}
\end{align}

\noindent
In \gls*{danse}, i.e.,~\eqref{eq:lmmseDANSE}, $\Gk[-k]^{i+1}\in\C[Q(K-1)][Q]$ is applied to $\zkd[-k]^i\in\C[Q(K-1)]$, while in \gls*{tidanse}, i.e.,~\eqref{eq:lmmseTIDANSE}, $\Gk^{i+1}\in\C[Q][Q]$ is applied to $\ettmk^i\in\C[Q]$.
All other nodes in $\Kmk$ do not update their filters and index $k$ is incremented at every iteration.
The target signal estimate at iteration $i$ at node $q\in\K$ is $\dhatk[q]^{i+1}= [\Wk[qq,\mathrm{D}]^{i+1,\Her} \:|\: \Gk[-q]^{i+1,\Her}]\ty[q,\mathrm{D}]$ in \gls*{danse} and $[\Wk[qq,\mathrm{TI}]^{i+1,\Her} \:|\: \Gk[q]^{i+1,\Her}]\ty[q,\mathrm{TI}]$ in \gls*{tidanse}.
The definitions are completed by the respective fusion rules: in \gls*{danse} $\Pk^{i+1} = \Wk[kk,\mathrm{D}]^{i+1}$ while in \gls*{tidanse} $\Pk^{i+1} = \Wk[k,\mathrm{TI}]^{i+1}(\Gk^{i+1})^{-1}$. Since other nodes do not update their filters, $\Pk[q]^{i+1}=\Pk[q]^{i}\fa q\in\Kmk$.

Both algorithms converge towards the centralized solution~\eqref{eq:mwfCentr} through iterations, although \gls*{tidanse} does so significantly more slowly than \gls*{danse} due to the lower number of \glspl*{dof} in $\ty[k,\mathrm{TI}]^i$ (i.e., $M_k + Q$) with respect to $\ty[k,\mathrm{D}]^i$ (i.e., $M_k + Q(K-1)$). In the following, we outline the \gls*{tidansep} algorithm, which enables faster convergence in any non-\gls*{fc} \glspl*{wasn} with a possibly time-varying topology.

%%%%%
\section{\gls*{tidansep}}\label{sec:TIDANSEplus}

In this section, we define \gls*{tidansep} at iteration $i$ with updating node $k$. Sequential node-updating is again considered, i.e., the index $k$ cycles over all nodes in a round-robin fashion.
% A \gls*{tidansep} iteration starts similarly to a \gls*{tidanse} iteration. 
The \gls*{wasn} is first pruned to a tree with root node $k$. Note that this tree may be different at each iteration, making \gls*{tidansep} robust to time-varying topologies. Each node $q\in\Kmk$ combines its $M_q$ sensor signals into a $Q$-dimensional fused signal $\zkd[q]^i$ via a fusion matrix $\Pk[q]^i\in\C[M_q][Q]$, to be defined later.
The fused signals are summed from the leaf nodes towards the root node $k$. We let $\Uk[q]^i$ denote the set of upstream neighbors of node $q$ and $\Ukb[q]^i$ the set of all nodes upstream of $q$ (i.e., $\Uk[q]^i\subseteq \Ukb[q]^i\fa q\in\K$) at iteration $i$. The sum $\ettkq[q][q']^i$ sent downstream by node $q$ to node $q'$ is:

\begin{equation}\label{eq:fusionflow}
    \ettkq[q][q']^i
    =  \zkd[q]^i + \sum_{m\in\Uk[q]^i}\ettkq[m][q]^i
    = \zkd[q]^i + \sum_{m\in\Ukb[q]^i}\zkd[m]^i.
\end{equation}

\noindent
Eventually, $|\Uk^i|$ individual partial in-network sums will reach $k$, stemming from each branch of the tree. The root node then constructs its observation vector as:

\vspace{-1em}

\begin{equation}\label{eq:yTildeRoot}
    \ty^i
    % = \Big[
    %     \yk^\T \: \underbrace{\ettkq[l_1][k]^{i\T}\:\dots\:\ettkq[l_B][k]^{i\T}}_{\ztk^{i\T}}
    % \Big]^\T
    = \begin{bmatrix}
        \yk^\T \: \ettkq[l_1][k]^{i\T}\:\dots\:\ettkq[l_B][k]^{i\T}
    \end{bmatrix}^\T
    = \begin{bmatrix}
        \yk^\T \:|\: \ztk^{i\T}
    \end{bmatrix}^\T,
\end{equation}

% \begin{equation}\label{eq:tilde_Mk}
%     \text{with}\:\:\tMk^i = M_k + Q|\Uk^i|,
% \end{equation}

\noindent
where $\{l_1,\dots,l_B\}=\Uk^i$ and the partial in-network sums $\{\ettkq[l][k]^i\}_{l\in\Uk^i}$ are stacked in $\ztk^i\in\C[{Q|\Uk^i|}]$.
Equation~\eqref{eq:yTildeRoot} clearly contrasts with the defintion of $\ty[k,\mathrm{TI}]^i$ for \gls*{tidanse}. The number of \glspl*{dof} at the updating node $k$ in \gls*{tidansep} is indeed $\tMk^i = M_k + Q|\Uk^i|$ instead of $M_k + Q$ in \gls*{tidanse}.
In a \gls*{fc} \gls*{wasn}, $\ty^i$ reduces to $\ty[k,\mathrm{D}]^i$ if the \gls*{fc} network is pruned to a star topology with $k$ as the hub.

The next step in \gls*{tidansep} is for the root node $k$ to estimate $\dk$ using $\ty^i$, computing $\tW^{i+1}\in\C[{\tMk^i}][Q]$ as:
% To estimate the node-specific target signal $\dk^i$ as a linear combination of the elements of $\ty^i$, the root node computes a new filter by solving the \gls*{lmmse} problem:
\vspace{-1em}

\begin{align}\label{eq:lmmseTIDANSEplus}
    \tW^{i+1}
    &= \underset{\Wkk,\Gtk}{\mathrm{arg\,min}}\:
    \E[
        {\left\|
            \dk - \begin{bmatrix}
                \Wkk^{\Her} \: \Gtk^\Her
            \end{bmatrix}\begin{bmatrix}
                \yk\\
                \ztk^i
            \end{bmatrix}
        \right\|^2_2}
    ]\\
    &=
    % \Big[
    %     \Wkk^{i+1,\T} \: \underbrace{\Gkq[k][l_1]^{i+1,\T}\:\dots\:\Gkq[k][l_B]^{i+1,\T}}_{\Gtk^{i+1,\T}}
    % \Big]^\T\label{eq:partition_tW_TIDANSEplus}\\
    \begin{bmatrix}
        \Wkk^{i+1,\T} \: \Gkq[k][l_1]^{i+1,\T}\:\dots\:\Gkq[k][l_B]^{i+1,\T}
    \end{bmatrix}^\T
    % =
    % \begin{bmatrix}
    %     \Wkk^{i+1,\T} \:|\: \Gtk^{i+1,\T}
    % \end{bmatrix}^\T
    \label{eq:partition_tW_TIDANSEplus}\\
    % &= \left(\Ryyt^i\right)^{-1}\Rydt^i\\
    &= \left(\Ryyt^i\right)^{-1}\Rsst^i\tEk.
    \label{eq:mwf_TIDANSEplus}
\end{align}

% \noindent
% This filter can be partitioned as:

% \begin{equation}\label{eq:tW_partition}
%   \tW^{i+1} =
%   \Big[
%     \Wkk^{i+1,\T} \: \underbrace{\Gkq[k][l_1]^{i+1,\T}\:\dots\:\Gkq[k][l_B]^{i+1,\T}}_{\Gtk^{i+1,\T}}
%   \Big]^\T,
% \end{equation}

% \noindent
% with $\Wkk^{i+1}\in\C[M_k][Q]$ the filter applied to the local sensor signals $\yk$ and $\Gtk^{i+1}\in\C[Q|\Uk^i|][Q]$ the filter applied to $\ztk^{i}$. The $\Gtk^{i+1}$ matrix is itself composed of $|\Uk^i|$ submatrices $\Gkq[k][l]^{i+1}\in\C[Q][Q],\:\forall\:l\in\Uk^i$, where $\Gkq[k][l]^{i+1}$ is applied to $\ettkq[l][k]^i$.

% Note that all nodes in the \gls*{wasn} have their own node-specific filter $\Wkk[q]^{i}$ and use it to compute their own fusion matrix $\Pk[q]^i$ (explicitly defined later). 

\noindent
Equation~\eqref{eq:partition_tW_TIDANSEplus} shows a partitioning of $\tW^{i+1}$ where $\Wkk^{i+1}\in\C[M_k][Q]$ is applied to $\yk$ and 
% $\Gtk^{i+1}\in\C[Q|\Uk^i|][Q]$ the filter applied to $\ztk^{i}$. The $\Gtk^{i+1}$ matrix is itself composed of $|\Uk^i|$ submatrices $\Gkq[k][l]^{i+1}\in\C[Q][Q]\fa l\in\Uk^i$, where
$\Gkq[k][l]^{i+1}\in\C[Q][Q]$ to $\ettkq[l][k]^i\fa l\in\Uk^i$. 
Equation~\eqref{eq:mwf_TIDANSEplus} shows the \gls*{mwf} solution to~\eqref{eq:lmmseTIDANSEplus} where $\Ryyt^i = \E[{\ty^i\ty^{i\Her}}]$,
% $\Rydt = \E[{\ty^i\dk^{i\Her}}]$,
$\Rsst = \E[{\ts^i\ts^{i\Her}}]$ (having split $\ty^i$ into a desired signal component $\ts^i$ and a noise component $\tn^i$), and $\tEk=[\Ekk^\T\:|\:\mathbf{0}]^\T$.
Non-root nodes do not update their filter, i.e., $\tW[q]^{i+1} = \tW[q]^{i}\fa q\in\Kmk$.
The root node target signal estimate is directly obtained as $\dhatk^{i+1} = \tW^{i+1,\Her}\ty^i$.

\begin{figure}[h]
    \centering
    \includegraphics[width=\columnwidth,trim={0 0 0 0},clip=false]{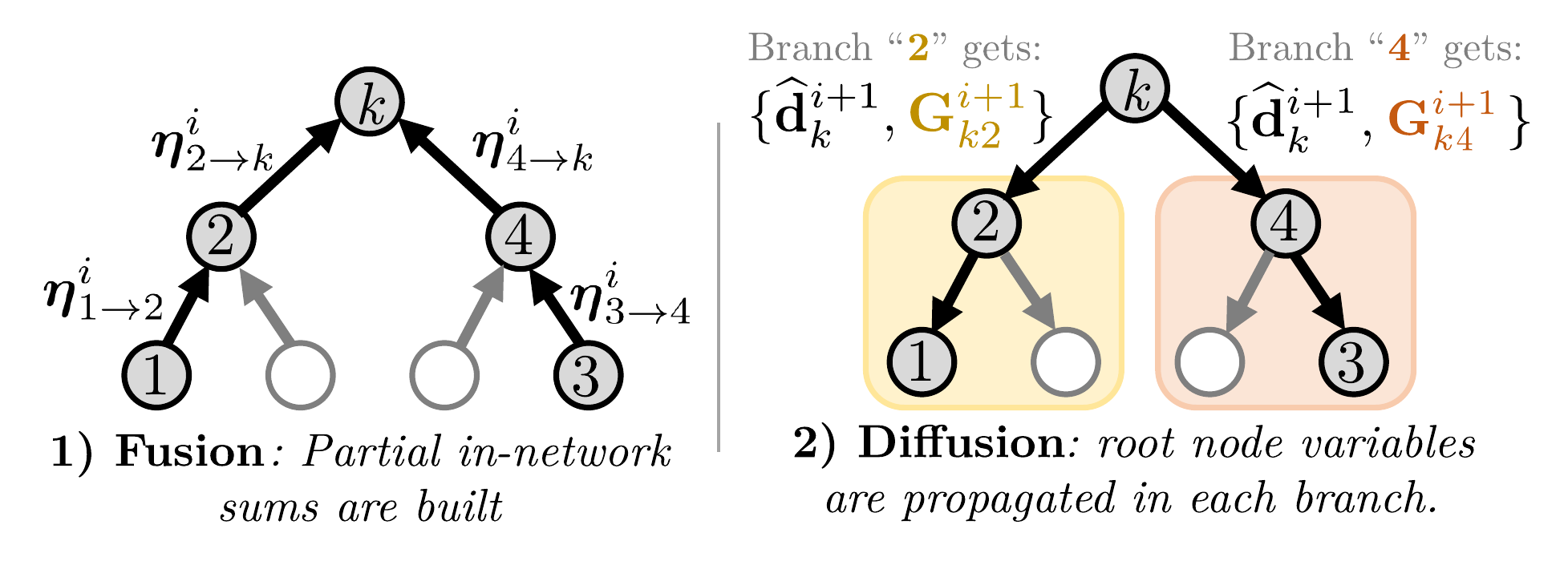}
    \caption{Schematic description of the \gls*{tidansep} data flows before and after the filter update at node $k$ (cf.~\eqref{eq:mwf_TIDANSEplus}).}
    \label{fig:fusiondiffusion}
\end{figure}

In order to estimate the target signal at non-updating nodes, the signal $\dhatk^{i+1}$ is propagated through the entire \gls*{wasn} and the matrix $\Gkq[k][l]^{i+1}$ is propagated through the tree branch stemming from node $l\fa l\in\Uk^i$. The \gls*{tidansep} data flows are summarized in~\figref{fig:fusiondiffusion}. Note that the transmit power required to propagate the matrices $\{\Gkq[k][l]^{i+1}\}_{l\in\Uk^i}$ is negligible compared to the necessary transmission of $\dhatk^{i+1}$. In other words, \gls*{tidansep} does not require significantly more transmit power than \gls*{tidanse}, in which the sum of all fused signals must be propagated through the \gls*{wasn} (cf.~\secref{subsec:danse_and_tidanse}). 

% In order to estimate the node-specific target signal $\dk[q]$ at any non-updating node $q\in\Kmk$, the following 
% For any $l\in\Uk^i$, the $Q\times Q$ filter $\Gkq[k][l]^{i+1}$ is applied to $\ettkq[l][k]^i$ and thus to all fused signals that were used to build it, namely $\{\zkd[q]^i\}_{q\in\Ukb[l]^i\cup l}$.
% \noindent
The fusion matrix at node $q$ is then defined as:
%  results in the sum of all fused signals being equal to the target signal estimate at the root node. The fusion matrix at iteration $i+1$ is thus defined as:

% The fusion matrices in~\eqref{eq:fused_signal_zk} at iteration $i+1$ are defined as:

\vspace{-1em}

\begin{align}\label{eq:PkDef}
    \Pk[q]^{i+1} &= \Wkk[q]^{i+1}\Tk[q]^{i+1}\:\:\text{where}\:\:\Tk[q]^{i+1} = \begin{cases}
        \Tk[q]^{i}\Gkq[k][n^i(q)]^{i+1}&\forall\:q\in\Kmk\\
        \mathbf{I}_Q&\text{for $q=k$}
    \end{cases},
\end{align}

\noindent
where $n^i(q)$ denotes the neighbor node of $k$ that belongs to the same branch as $q$, i.e., $n^i(q)\in\Uk^i$ and $q\in\Ukb[n^i(q)]^i$.
Remarkably, using this fusion matrix definition, the target signal estimate at the root node $\dhatk^{i+1}$ is equal to the sum of fused signals at the end of iteration $i$ (i.e., once the fusion matrices have been updated using~\eqref{eq:PkDef}) since:

\vspace{-1em}

\begin{align}
    \dhatk^{i+1}
    % &=
    % \tW^{i+1,\Her}\ty^i\\
    &=
    \Wkk^{i+1,\Her}\yk + \sum_{l\in\Uk^i}\Gkq[k][l]^{i+1,\Her}\Bigg(
        \zkd[l]^i + \sum_{m\in\Ukb[l]^i}\zkd[m]^i
    \Bigg)\label{eq:rewrite_dk_eq3}\\
    &= \Pk^{i+1,\Her}\yk + \sum_{q\in\Kmk}\Gkq[k][n^i(q)]^{i+1,\Her}\Pk[q]^{i\Her}\yk[q]\label{eq:rewrite_dk_eq4}\\
    &= \sum_{q\in\K}\Pk[q]^{i+1,\Her}\yk[q].\label{eq:dansep_dhat_expanded}
\end{align}

\noindent
One way to understand~\eqref{eq:dansep_dhat_expanded} is to observe that~\eqref{eq:PkDef} essentially orients the fusion matrices of all the nodes towards the signal estimation objective of node $k$. Combining this observation with~\eqref{eq:centr_same_sol_space}, it becomes apparent that $\dhatk[q]^{i+1}$ is equal to $\dhatk^{i+1}$ up to a $Q\times Q$ transformation. In \gls*{tidansep}, the target signal estimate at any node can then in fact be obtained as:

\vspace{-1em}
\begin{align}\label{eq:desSigEst}
    \dhatk[q]^{i+1} &= \left(\Tk[q]^{i+1,\Her}\right)^{-1}\dhatk^{i+1},\:\forall\:q\in\K,
\end{align}

\noindent
which completes the definition of \gls*{tidansep}, as summarized in~\algref{alg:tidansep}. 
It is noted that the network-wide \gls*{tidansep} filter at node $q\in\K$ (such that $\Wk[q]^{i+1,\Her}\mathbf{y} = \dhatk[q]^{i+1}$) can be written based on~\eqref{eq:PkDef} and~\eqref{eq:desSigEst} as:

\begin{equation}\label{eq:nw_filter_dansep}
    \Wk[q]^{i+1} =
    \begin{bmatrix}
        \Pk[1]^{i+1}\left(\Tk[q]^{i+1}\right)^{-1}\\
        \vdots\\
        \Wkk[q]^{i+1}\\
        \vdots\\
        \Pk[K]^{i+1}\left(\Tk[q]^{i+1}\right)^{-1}\\
    \end{bmatrix}
    =
    \begin{bmatrix}
        \Wkk[1]^{i+1}\Tk[1]^{i+1}\left(\Tk[q]^{i+1}\right)^{-1}\\
        \vdots\\
        \Wkk[q]^{i+1}\\
        \vdots\\
        \Wkk[K]^{i+1}\Tk[K]^{i+1}\left(\Tk[q]^{i+1}\right)^{-1}\\
    \end{bmatrix},
\end{equation}

\noindent
which will be used in~\secref{sec:numericalExperiments}.

In a \gls*{fc} \gls*{wasn}, \gls*{tidansep} offers an advantage in terms of transmit power over \gls*{danse} since, in \gls*{tidansep}, non-updating nodes only require $\dhatk^{i+1}$ to compute their own target signal estimate $\dhatk[q]^{i+1}$ via~\eqref{eq:desSigEst}. In \gls*{danse}, however, node $q$ must construct $\ty[q,\mathrm{D}]^i$ to obtain $\dhatk[q]^{i+1}$, which is only possible if every $q'\in\Kmk[q]$ transmits its own $\zkd[q']^i$.
% \textit{Theorem 1:} If $\dk[q] = \Ekk[q]^\T\Psibk[q]\mathbf{s}^\mathrm{lat}$ with $\Ekk[q]^\T\Psibk[q]$ full-rank$\fa q\in\K$, then the solution space defined by~\eqref{eq:nw_filter_dansep} includes the centralized optimal filters $\hWk[q]$ as in~\eqref{eq:mwfCentr}$\fa q\in\K$.
% \textit{Proof:} 

\begin{algorithm}%[h]
    \caption{The \gls*{tidansep} algorithm}\label{alg:tidansep}
    \begin{algorithmic}[1]
        \STATE $k \gets 1$.
        \STATE Initialize $\Wkk[q]^0$ and $\Pk[q]^0$ randomly$\fa q\in\K$.
        \FOR{$i=1,2,3,\dots$}
            \STATE Prune \gls*{wasn} to tree with root node $k$.
            \STATE Each node $q\in\K$ computes $\zkd[q]^i = \Pk[q]^{i\Her}\yk[q]$.
            \STATE Partial in-network sums $\{\ettkq[l][k]\}_{l\in\Uk^i}$ are built via~\eqref{eq:fusionflow} and made available at root node $k$.
            \STATE Root node $k$ builds $\ty^i$ as in~\eqref{eq:yTildeRoot}.
            \STATE Root node $k$ computes $\tW^{i+1}$ as in~\eqref{eq:mwf_TIDANSEplus}.\label{algstep:Wtilde}
            \STATE Root node $k$ computes $\dhatk^{i+1}=\tW^{i+1,\Her}\ty^i$.
            \STATE $\dhatk^{i+1}$ is propagated through the \gls*{wasn} and $\Gkq[k][l]^{i+1}$ through the branch stemming from $l\fa l\in\Uk^i$.
            \STATE Each node $q\in\K$ computes $\Pk[q]^{i+1}$ via~\eqref{eq:PkDef}.
            \STATE Each node $q\in\Kmk$ computes $\dhatk[q]^{i+1}$ via~\eqref{eq:desSigEst}.
            \STATE $k \gets (k+1)\mod K$.
        \ENDFOR
    \end{algorithmic}
\end{algorithm}

% \textit{Theorem:} In any connected topology-unconstrained \gls*{wasn}, if $\dk[q] = \Psibkov[q]\mathbf{s}^\mathrm{lat}$ with $\Psibkov[q]$ full-rank$\fa q\in\K$, then \gls*{tidansep} converges in node $q$ to the centralized optimal filter $\hWk[q]$ in~\eqref{eq:mwfCentr}$\fa q\in\K$.

% \textit{Proof:} Omitted.

%%%%%
\section{Tree-pruning}\label{sec:treepruning}

The definition of \gls*{tidansep} does not impose any constraint on the tree-pruning strategy. The algorithm can be applied as long as the \gls*{wasn} is pruned to a tree (any tree) at the beginning of every iteration.
Although, when considering \gls*{wasn} tree-pruning, \glspl*{mst} are typically preferred as they minimize the energy required to transfer data through the tree, a different strategy may be selected to increase the convergence speed of \gls*{tidansep}.%, as justified in the following.

% \todoi{Add citation for ``popular option'' $\uparrow$?}
% \todoi{Talk about peer-to-peer clearly}

Indeed, the convergence speed of any iterative distributed algorithm that relies on \gls*{lmmse}-based updates as in~\eqref{eq:lmmseDANSE},~\eqref{eq:lmmseTIDANSE}, or~\eqref{eq:lmmseTIDANSEplus} increases with the number of available \glspl*{dof} used when solving the \gls*{lmmse} problem. This phenomenon leads to the slower convergence of \gls*{tidanse} since, there, each \gls*{lmmse} problem only has $M_k+Q$ \glspl*{dof} compared to $M_k + Q(K-1)$ in \gls*{danse} in a \gls*{fc} \gls*{wasn}. Conversely, the size of the observation vector at the updating node in \gls*{tidansep}~\eqref{eq:yTildeRoot} is $M_k + Q|\Uk^i|$. Therefore, increasing $|\Uk^i|$ can be expected to increase the \gls*{tidansep} convergence speed.

\begin{figure}[h]
    \centering
    \includegraphics[width=\columnwidth,trim={0 0 0 0},clip=false]{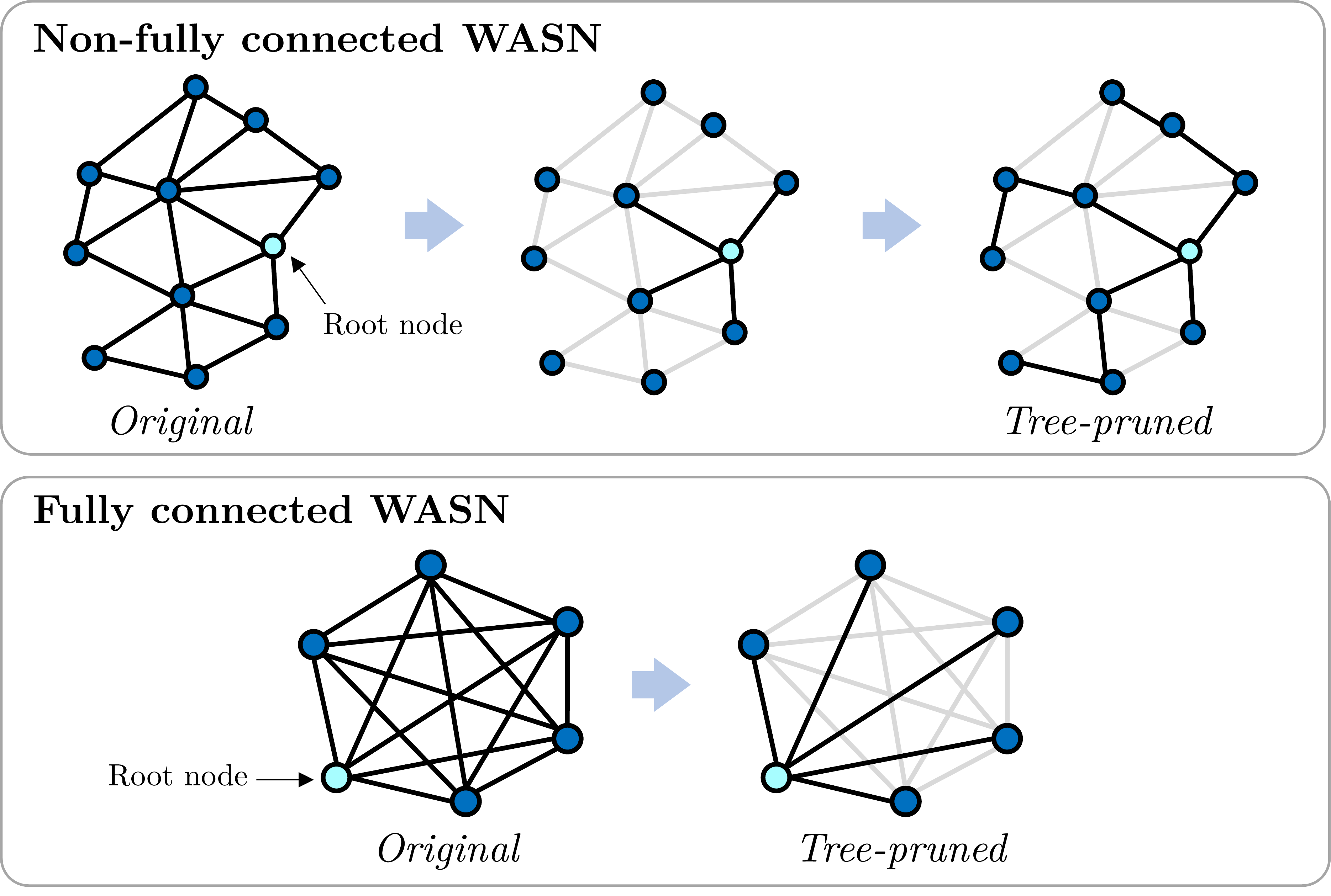}
    \caption{\gls*{mmut} tree-pruning applied to a non-\gls*{fc} \gls*{wasn} and a \gls*{fc} \gls*{wasn}.}
    \label{fig:examples_pruning}
\end{figure}

% The largest number of connections that the root node $k$ can have after pruning is the number of neighbors that $k$ has in the original non-pruned \gls*{wasn}. 
We thus propose to prune any (\gls*{fc} or not) \gls*{wasn} in which \gls*{tidansep} is deployed to a tree with the largest $|\Uk^i|$ possible$\fa i$. To achieve this, we simply ensure that all existing connections to updating node $k$ are preserved after pruning.
% , and with a minimum total edge cost which may be defined, e.g., as the Euclidean distance between nodes. This is essentially a constrained version of the Kruskal's \gls*{mst}-finding algorithm~\cite{kruskalShortest} and represents a trade-off between \gls*{tidansep} convergence speed and communication cost.
This strategy is referred to as \gls*{mmut} pruning. % is defined in~\algref{alg:mmut}.
% Since $k$ is the root node, each connection then corresponds to a branch.
The remaining part of the \gls*{wasn} can be pruned randomly or according to a criterion. In this paper, we choose to minimize the distance between connected nodes in the resulting tree.
Examples for a non-\gls*{fc} \gls*{wasn} and a \gls*{fc} \gls*{wasn} are given in~\figref{fig:examples_pruning}, showing that \gls*{mmut} pruning in a \gls*{fc} \gls*{wasn} results in a star topology. It should be noted that transferring data through a \gls*{mmut}-pruned tree might require slightly more transmit power than doing so through a \gls*{mst}. %However, the \gls*{wasn}'s transmit power budget can still be dynamically maintained by selecting an appropriate tree-pruning strategy at each \gls*{tidansep} iteration. % This trade-off between convergence speed and communication cost is important to consider when designing solutions for particular scenarios.

% \begin{algorithm}%[h]
%     \caption{\gls*{mmut} pruning}\label{alg:mmut}
%     \begin{algorithmic}[1]
%         \STATE Let $\G$ denote the graph corresponding to a non-pruned topology-unconstrained \gls*{wasn} and $k$ the root node.
%         \STATE Initialize tree $\Tree(k)$ with all edges of $\G$ connected to $k$.\label{algstep:mmut_step2}
%         \STATE Sort the remaining edges by increasing weight (e.g., Euclidean distance between the two corresponding nodes) and list them accordingly in $\mathbf{e}_{\backslash k}$.
%         \FOR{$e$ in $\mathbf{e}_{\backslash k}$}
%         \IF{$e\notin \Tree(k)$ \textbf{and} there is no path in $\Tree(k)$ between the nodes linked by $e$}
%             \STATE Add $e$ to $\Tree(k)$.
%         \ENDIF
%         \ENDFOR\label{algstep:mmut_stepFinal}
%     \end{algorithmic}
% \end{algorithm}

\vspace{-.5em}

%%%%%
\section{Simulations}\label{sec:numericalExperiments}

The convergence properties of \gls*{tidansep} are showcased in this section via simulations. The code used to obtain the results is available online\footnote{\url{https://github.com/p-didier/tidanseplus_batch} (Accessed: 10 March 2025)}. Static \gls*{wasn} topologies are considered, where the communication links remain fixed.
% It is noted that \gls*{tidansep} can function in time-varying \gls*{wasn} topologies, although this is not considered in these simulations.
All results are averages over 10 randomly generated sensing environments, where each environment consists of a \gls*{wasn} with $K{=}10$ nodes and $M_q{=}3$ sensors per node, $\forall\:q\in\K$, $Q{=}1$ desired signal source, and three noise sources. %One-dimensional signals are exchanged between nodes, i.e., $Q{=}1$.
The \gls*{wasn} is generated by randomly placing the nodes and sources on a 2-D plane in a $5\,\text{m}\times 5\,\text{m}\times 5$\,m 3-D space, ensuring a minimum distance of 0.5\,m between sources and nodes. The sensors of each node are randomly placed in a disc of 10\,cm radius around the node position. A communication radius is initialized at 1.5\,m and increased until the \gls*{wasn} is connected. % The desired signal source and noise sources are randomly placed ensuring a minimum distance of 0.5\,m with all sensors.

We simulate frequency-domain processing at an abritrarily chosen frequency.
For simplicity, the 
% signal $s^\mathrm{lat}$ and $\mathbf{n}^\mathrm{lat}$ are drawn from the uniform distribution over [-0.5, 0.5] and the
steering matrix entry between any source and sensor of any node is set as the 3-D free-field Green's function with a speed of sound of $c=343$ m/s, which allows to represent delays and amplitude differences between sensors. The \glspl*{snr} obtained at the first sensor of each node across all simulations range from -17.7\,dB to 5.9\,dB.

% \todoi{``$1/d_{sm}$ is a bit too simple, not?''}

To unambiguously assess the convergence properties, the 
\glspl*{scm} required to compute the \glspl*{mwf} are obtained via oracle knowledge of the steering matrices, as $\Rss=\Psibk[]\mathbf{R}_\mathbf{ss}^\mathrm{lat}\Psibk[]^\Her$ where $\Psibk[] = [\Psibk[1]^\T\:\dots\:\Psibk[K]^\T]^\T$ and $\mathbf{R}_\mathbf{ss}^\mathrm{lat}\in\C[Q][Q]$ is a diagonal matrix with the powers of the latent desired signals in $\mathbf{s}^\mathrm{lat}$ on the diagonal. The \gls*{tidansep} \gls*{scm} $\Rsst^i$ is obtained as $\Ck^{i\Her}\Rss\Ck^i$, for any $k$, where $\mathbf{C}_k^i\in\C[M][{\tMk^i}]$ contains an appropriate arrangement of the fusion matrices $\{\Pk[q]^i\}_{q\in\Kmk}$. The \glspl*{scm} $\Ryy$ and $\Ryyt^i$ are computed similarly. Actual microphone signals are not computed in these simulations.

% \todoi{Add info about the SNR, the self-noise power, source localizations}

The convergence speed of \gls*{tidansep} is compared to that of \gls*{tidanse} and \gls*{danse} in terms of \gls*{mse} between the network-wide filters\footnote{The \gls*{tidansep} network-wide filter expression is given in~\eqref{eq:nw_filter_dansep}, in \cite[pp.8]{bertrandDistributed} for \gls*{danse} filters, and in~\cite[pp.6]{szurleyTopology} for \gls*{tidanse} filters.} $\Wk[q]^{i}$ at iteration $i$ and the centralized filters $\hWk[q]$ from~\eqref{eq:mwfCentr}, as:

\begin{equation}\label{eq:msew}
    \mathrm{MSE}_{W}^i = \tfrac{1}{K} \textstyle\sum_{q=1}^K \|
        \Wk[q]^{i} - \hWk[q]
    \|^2_F.
\end{equation}

\noindent
where $\|\cdot\|_F$ denotes the Frobenius norm.
The magnitude of $\mathrm{MSE}_{W}^i$ can significantly vary across sensing environments. In order to unambiguously depict the average $\mathrm{MSE}_{W}^i$ obtained across sensing environments on a logarithmic axis, we compute a geometric mean $\overline{\mathrm{MSE}}_{W}^i = (\prod_{m=1}^{N_\mathrm{SE}} \mathrm{MSE}_{W}^i(m))^{1/N_\mathrm{SE}}$ for each algorithm, where $N_\mathrm{SE}$ denotes the number of sensing environments and $\mathrm{MSE}_{W}^i(m)$ the value of~\eqref{eq:msew} obtained for the $m$-th sensing environment.

As discussed in~\secref{sec:treepruning}, the number of neighbors for the updating node (i.e., $|\Uk^i|$) is expected to influence the convergence speed of \gls*{tidansep}. Consider any \gls*{wasn} with adjacency matrix $\mathbf{\Lambda}\in\{0,1\}^{K\times K}$, i.e., $[\mathbf{\Lambda}]_{q,l} = 1$ if $q$ and $l\neq q$ are connected, 0 otherwise, and $[\mathbf{\Lambda}]_{q,q} = 0\fa q\in\K$. We define the following metric to quantify the connectivity of the \gls*{wasn}:

\begin{equation}\label{eq:connectivity}
    % C = \frac{\mathbf{1}^\T\mathbf{\Lambda}\mathbf{1} - 2K}{K(K-3)},
    C = (\mathbf{1}^\T\mathbf{\Lambda}\mathbf{1} - 2K)/(K(K-3)),
\end{equation}

\noindent
where $\mathbf{1}$ denotes the $K$-dimensional all-ones column vector. In a \gls*{fc} \gls*{wasn}, $C=1$. In a tree \gls*{wasn}, $C=0$. %As $C$ is unambiguously bounded between 0 and 1, we prefer it over the algebraic connectivity.
We evaluate \gls*{tidansep} in \glspl*{wasn} with various values of $C$ by randomly removing or establishing communication links between nodes of the original \gls*{wasn} (ensuring that the network remains connected) until a desired $C$ is obtained. % This process is repeated for different values of $C$.

\begin{figure}[h]
    \centering
    \includegraphics[width=\columnwidth,trim={0 0 0 0},clip=false]{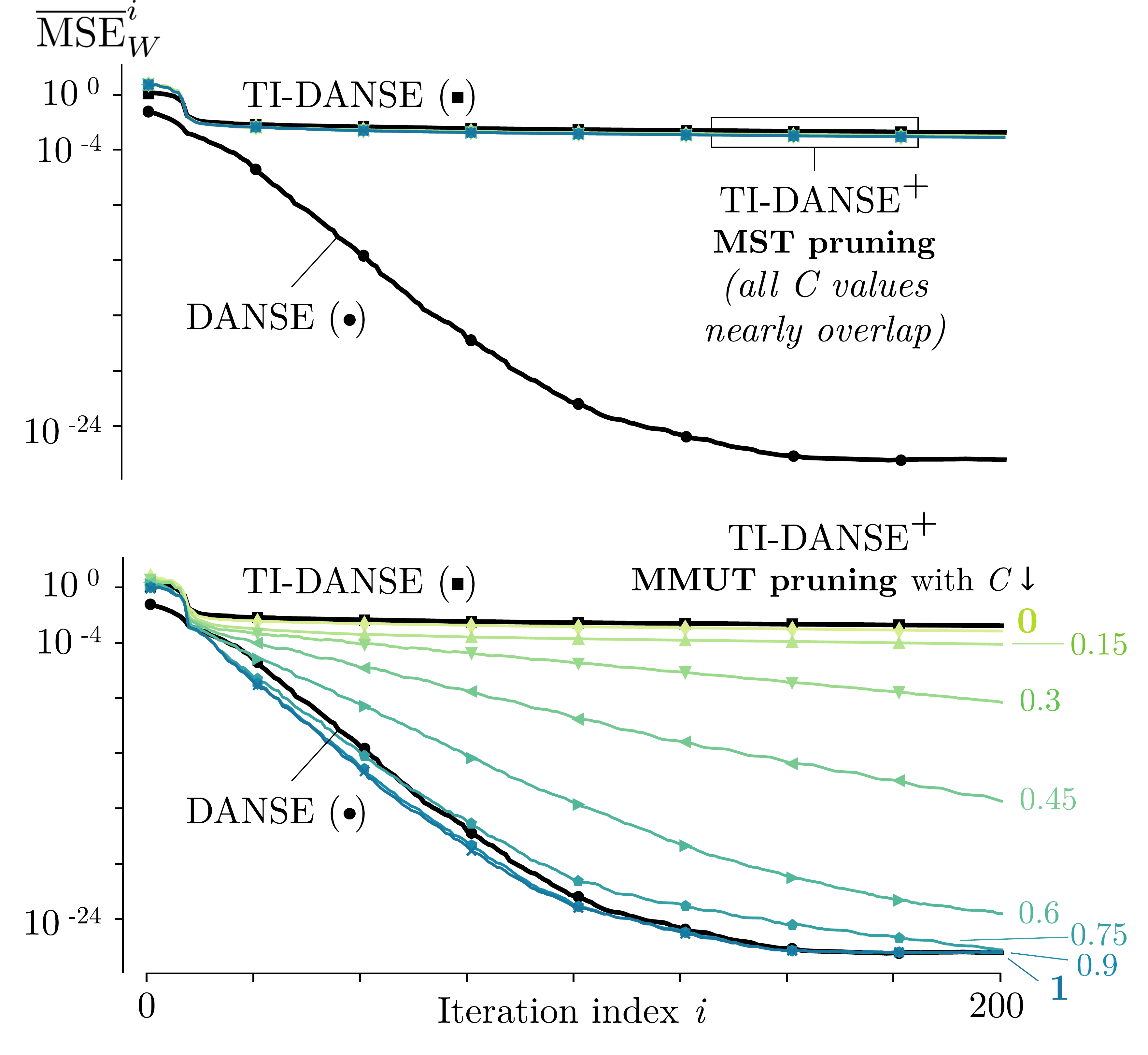}
    \caption{$\overline{\mathrm{MSE}}_{W}^i$ for \gls*{tidansep} with different values of connectivity $C$, with comparison to \gls*{tidanse} and \gls*{danse}. The results using \gls*{mst} or \gls*{mmut} pruning are depicted in the top and bottom plots, respectively, highlighting the benefit of \gls*{mmut} pruning for fast \gls*{tidansep} convergence.}
    \label{fig:res_msew}
\end{figure}

The metric $\overline{\mathrm{MSE}}_{W}^i$ is computed when pruning the \gls*{wasn} at each iteration $i$ to either an \gls*{mst} using Kruskal's algorithm~\cite{kruskalShortest} or a tree obtained via \gls*{mmut} pruning. The results for \gls*{danse} are obtained as if the \gls*{wasn} would be \gls*{fc}. The results for \gls*{tidanse} are independent of $C$ since nodes use the global in-network sum. All results are shown in~\figref{fig:res_msew}.

% \begin{figure}[h]
%     \centering
%     \begin{scriptsize}
%         \input{pp_conn_msew_minst_geomean.tex}
%         \input{pp_conn_msew_mmut_geomean.tex}
%     \end{scriptsize}
%     \caption{$\overline{\mathrm{MSE}}_{W}^i$ for \gls*{tidansep} with different values of connectivity $C$, with comparison to \gls*{tidanse} and \gls*{danse}, the latter obtained as if the \gls*{wasn} was \gls*{fc}. Averages over 10 randomly generated sensing environments. The results using \gls*{mst} or \gls*{mmut} pruning are depicted in the top and bottom plots, respectively.}
%     \label{fig:res_msew}
% \end{figure}

The results showcase the slow convergence of \gls*{tidanse} compared to that of \gls*{danse} in an equivalent \gls*{fc} \gls*{wasn}. The importance of the tree-pruning strategy is apparent for \gls*{tidansep}. When pruning the \gls*{wasn} to an \gls*{mst}, the original connectivity $C$ does not play any significant role. However, when using \gls*{mmut} pruning, the \gls*{tidansep} convergence speed increases with the connectivity in the original unpruned \gls*{wasn}. In fact, for $C{=}1$ (i.e., in a \gls*{fc} \gls*{wasn}) with \gls*{mmut} pruning, \gls*{tidansep} converges as fast as \gls*{danse}. The benefit of using \gls*{tidansep} instead of \gls*{tidanse} is also clearly visible for $0<C<1$, even in sparse networks with few inter-node connections. %Even with $C{=}0$ (i.e., in \glspl*{wasn} with a tree topology) or when pruning to an \gls*{mst}, the \gls*{tidansep} algorithm reaches a lower $\overline{\mathrm{MSE}}_{W}^i$ in fewer iterations than \gls*{tidanse}. This is due to the fact that, even in a tree topology with $K>2$, at least one node has more than one neighbor and can, through \gls*{tidansep}, exploit several partial in-network sums individually to update its filters. 

%%%%%
\section{Conclusions}\label{sec:conclusion}

In this paper, we have introduced \gls*{tidansep}, a topology-independent distributed signal estimation algorithm that addresses the slow convergence of the \gls*{tidanse} algorithm in \glspl*{wasn} and improves transmit power efficiency in \gls*{fc} \glspl*{wasn} compared to the \gls*{danse} algorithm by requiring the transmission of fewer signals. The convergence speed increase in \gls*{tidansep} is achieved by letting the updating node exploit each partial in-network sum of fused signals as received from its neighbors as a separate \gls*{dof} in its local estimation problem.
Our simulated experiments have shown that an appropriate choice of tree-pruning strategy can make \gls*{tidansep} significantly outperform \gls*{tidanse} in terms of convergence speed, even in non-\gls*{fc} \glspl*{wasn} with relatively few inter-node connections. In future work we aim at formally proving the convergence and optimality of \gls*{tidansep} and consider generalized eigenvalue decomposition-based solutions to the local \gls*{lmmse} problems. %All in all, when combined to an appropriate tree-pruning strategy, \gls*{tidansep} is an algorithmic solution that improves on \gls*{tidanse} by allowing faster convergence.

\end{document}

%% file: acro.tex
\newacronym{wasn}{WASN}{wireless acoustic sensor network}
\newacronym{danse}{DANSE}{distributed adaptive node-specific signal estimation}
\newacronym{tidansep}{TI-DANSE$^+$}{topology-independent distributed adaptive node-specific signal estimation PLUS[TO BE CHANGED]}
\newacronym{gevddansep}{TI-GEVD-DANSE$^+$}{topology-independent GEVD-based distributed adaptive node-specific signal estimation PLUS[TO BE CHANGED]}
\newacronym{gevd_or_dansep}{TI-(GEVD-)DANSE$^+$}{topology-independent (GEVD-based) distributed adaptive node-specific signal estimation PLUS[TO BE CHANGED]}
\newacronym{tdanse}{T-DANSE}{tree-DANSE}
\newacronym{tidanse}{TI-DANSE}{topology-independent distributed adaptive node-specific signal estimation}
\newacronym{tigevddanse}{TI-GEVD-DANSE}{topology-independent GEVD-based distributed adaptive node-specific signal estimation}
\newacronym{ti_or_gevddanse}{TI-(GEVD-)DANSE}{topology-independent (GEVD-based) distributed adaptive node-specific signal estimation}
\newacronym{gevddanse}{GEVD-DANSE}{GEVD-based distributed adaptive node-specific signal estimation}
\newacronym{gevd}{GEVD}{generalized eigenvalue decomposition}
\newacronym{lmmse}{LMMSE}{linear minimum mean square error}
\newacronym{mmse}{MMSE}{minimum mean square error}
\newacronym{mwf}{MWF}{multichannel Wiener filter}
\newacronym{gevdmwf}{GEVD-MWF}{GEVD-based multichannel Wiener filter}
\newacronym{scm}{SCM}{spatial covariance matrix}
\newacronym{snr}{SNR}{signal-to-noise ratio}
\newacronym{mse}{MSE}{mean square error}
\newacronym{vad}{VAD}{voice activity detector}
\newacronym{mst}{MST}{minimum spanning tree}
\newacronym{gsbcd}{GSBCD}{Gauss-Seidel block-coordinate descent}
\newacronym{ao}{AO}{alternating optimization}
\newacronym{mmut}{MMUT}{multiple max-$|\Uk|$ trees}
\newacronym{mc}{MC}{Monte-Carlo}
\newacronym{ac}{AC}{algebraic connectivity}
\newacronym{fc}{FC}{fully connected}
\newacronym{stft}{STFT}{short-term Fourier transform}
\newacronym{dft}{DFT}{discrete Fourier transform}
\newacronym{dof}{DoF}{degree of freedom}